\documentstyle[12pt]{article}
\catcode`\@=11

\def\@normalsize{\@setsize\normalsize{15pt}\xiipt\@xiipt
\abovedisplayskip 14pt plus3pt minus3pt%
\belowdisplayskip \abovedisplayskip
\abovedisplayshortskip  \z@ plus3pt%
\belowdisplayshortskip  7pt plus3.5pt minus0pt}

\def\small{\@setsize\small{13.6pt}\xipt\@xipt
\abovedisplayskip 16pt plus3pt minus3pt%
\belowdisplayskip \abovedisplayskip
\abovedisplayshortskip  \z@ plus3pt%
\belowdisplayshortskip  7pt plus3.5pt minus0pt
\def\@listi{\parsep 4.5pt plus 2pt minus 1pt
            \itemsep \parsep
            \topsep 9pt plus 3pt minus 3pt}}

\def\underline#1{\relax\ifmmode\@@underline#1\else
        $\@@underline{\hbox{#1}}$\relax\fi}
\@twosidetrue

\catcode`@=12

\catcode`\@=11

\def\thesection{\Roman{section}}

\def\ps@headings{\def\@oddfoot{}\def\@evenfoot{}}
\def\@oddhead{\hbox{}\hfill
        \makebox[.5\textwidth]{\raggedright\ignorespaces --\thepage{}--
        \hfill }}
\def\@oddhead{\hbox{}\hfill --\thepage{}-- \hfill}
\def\@evenhead{\@oddhead}

\ps@headings
\catcode`\@=12

\relax

\newcounter{appendix}
\def\appendix{\par
 \reset{equation}
 \addtocounter{appendix}{1}
 \def\thesection{\Alph{appendix}.}
 \def\ksection{\Alph{appendix}}}

\begin{document}
\newcommand{\dal}{\raisebox{0.085cm}
{\fbox{\rule{0cm}{0.07cm}\,}}}
\newcommand{\eg}{{\em e.g.~}}
\newcommand{\etc}{{\em etc.}}
\newcommand{\cf}{{\em c.f.~}}
\newcommand{\al}{\alpha^{\prime}}
\newcommand{\mst}{M_{\scriptscriptstyle \!S}}
\newcommand{\mpl}{M_{\scriptscriptstyle \!P}}
\newcommand{\half}{\textstyle\frac{1}{2}}
\newcommand{\shalf}{{\textstyle\frac{1}{\sqrt{2}}}}
\newcommand{\dv}{\int{\rm d}^4x\sqrt{g}}
\newcommand{\act}{\widetilde{\Gamma}}
\newcommand{\lv}{\left\langle}
\newcommand{\rv}{\right\rangle}
\newcommand{\ph}{\varphi}
\newcommand{\sbar}{\,\overline{\! S}}
\newcommand{\tbar}{\overline{T}}
\newcommand{\sigmab}{\bar{\sigma}}
\newcommand{\lambdab}{\bar{\lambda}}
\newcommand{\psib}{\bar{\psi}}
\newcommand{\chib}{\bar{\chi}}
\newcommand{\zetab}{\bar{\zeta}}
\newcommand{\ybar}{\overline{Y}}
\newcommand{\phb}{\overline{\varphi}}
\newcommand{\cm}{Commun.\ Math.\ Phys.~}
\newcommand{\pr}{Phys.\ Rev.\ D~}
\newcommand{\pl}{Phys.\ Lett.\ B~}
\newcommand{\np}{Nucl.\ Phys.\ B~}
\newcommand{\e}{{\rm e}}
\newcommand{\gsi}{\,\raisebox{-0.13cm}{$\stackrel{\textstyle
>}{\textstyle\sim}$}\,}
\newcommand{\lsi}{\,\raisebox{-0.13cm}{$\stackrel{\textstyle
<}{\textstyle\sim}$}\,}
\begin{titlepage}
\begin{flushright}
\vskip -1.0cm
NUB--3064, PUPT-1400, SNUTP-93-29\\
April 1993
\end{flushright}
\vskip -0.2in
\begin{center}
\large
\sc Instanton Effects in Supergravity Theories$^{\star}$
\end{center}\par \noindent
\vskip-1cm
\begin{center}
     {\large Soo-Jong Rey}
  \par \vskip -.1in \noindent
  {\sl Joseph Henry Laboratory, Princeton University,
 Princeton, NJ 08544, U.S.A.}\\[-2mm]
  {\sl Center for Theoretical Physics, Seoul National University,
 Seoul 151-742, KOREA}
  \end{center}
  \vskip-1cm
\begin{center}
      {\large  T.R. Taylor}
  \par \vskip -.1in \noindent
{\sl Department of Physics, Northeastern University,
Boston, MA 02115, U.S.A.}\\[0.4in]
\end{center}
\par \vskip -.30in
\begin{center}
{Abstract}\\
\end{center}
\begin{quote}
\vskip-0.3cm
Non-perturbative effects of instanton-like solutions are studied
within the framework of supergravity theories with field-dependent
gauge functions. Fermionic zero modes are constructed and
some typical correlation functions are evaluated.
The effects of instantons are very similar to those in globally
supersymmetric theories: they preserve supersymmetry
while breaking a chiral $U(1)$ symmetry. Non-perturbative amplitudes
receive corrections which are suppressed at large distances.
\end{quote}
\begin{flushleft}
\rule{5.1 in}{.007 in}\\[-4mm]
$^{\star}${\small Work supported in part by the Northeastern University
\vspace{-4mm} Research and Development Fund, in part by the \vspace{-4mm}
U.S. NSF under grant PHY--91--07809, in part by
the U.S. DOE under grant
DE-FG02-91ER40671, and in part by the KOSEF.}
\end{flushleft}
\end{titlepage}

Instanton solutions of Euclidean field equations exist in globally
supersymmetric gauge theories \cite{a,inst}
as well as in some more complicated theories which appear
as low-energy limits of superstring compactifications \cite{chs,rey}.
Non-perturbative effects of supersymmetric instantons have been
studied mainly within the framework of {\em global\/} supersymmetry
\cite{a,inst}.
It has been also asserted that some similar non-perturbative effects
are important in supergravity theories, particularly in the context
of the gaugino condensation mechanism of supersymmetry
breaking \cite{fgn}.
More recently, gaugino condensation has attracted considerable
attention in phenomenological applications of superstring theory \cite{din}.
The reason is that the
two basic ingredients of this supersymmetry breaking mechanism have a
very natural basis in superstring theory:
1) gauge groups with ``hidden'' gauginos are typical for
heterotic superstring compactifications,
2) gauge couplings are functions of neutral scalar fields \eg of the dilaton
and of the moduli.

In this letter, we study non-perturbative effects of instantons in
supergravity theories with field-dependent gauge functions.
Our main motivation has been to elaborate on the relation
between instantons and gaugino condensation in superstring-type
supergravity theories. This is the reason why we
focus our attention on a specific supergravity Lagrangian,
with the gauge coupling constant
determined by the vacuum expectation value (VEV)
of the dilaton field. We should mention already at this
point that our strategy can be readily applied to some
more complicated cases, as discussed at the end of this letter.

Our starting point is the supergravity Lagrangian describing
a general four-dimen\-sio\-nal heterotic superstring compactification
at the string tree level. In the standard supergravity formalism \cite{c},
it is specified by: 1) the gauge function $f=kS$, where $S$ is the
chiral dilaton superfield and $k$ is the Ka\v{c}-Moody level of
the algebra that generates the gauge group,
2) the K\"ahler potential $K=-\ln (S+\sbar)+G$, and 3) the superpotential $W$.
The functions $G$ and $W$ are model-dependent.
However, their forms are not important for the following considerations,
except for the fact that both these functions do not depend on the dilaton.
Since we are interested in classical solutions with
$W=0$, hereafter we neglect the superpotential.

In order to analyze classical field equations, it is convenient
to represent the dilaton and its supersymmetric partners
by a linear multiplet. The linear dilaton multiplet $L$ contains
the scalar dilaton $l$, the dilatino $\chi$, and the vector $h_0^m$
which is dual to the field-strength of the antisymmetric tensor field
$b_{pq}$: $h_0^m=\frac{1}{2}\epsilon^{mnpq}\partial_nb_{pq}$. In the
linear formulation, the full Lagrangian, including gauge kinetic terms,
originates from the d-density of the function \cite{lin}
${\cal L}=  (\widehat{L}/2)^{-1/2}\,  e^{-\al G/2}$,
with $\widehat{L}=L-2\al k\Omega$, where $\Omega$ is the
Chern-Simons superfield and $\al$ is a constant of length dimension 2.
In the context of string theory, $\al$ corresponds to the string
scale. The component form of this Lagrangian has been written explicitly
in \cite{adam}. In particular, one can see that the Kalb-Ramond field
is always contained in the vector
\begin{equation} \textstyle
h^m = h_0^m-\frac{\al k}{2}\omega^m,
\end{equation}
where $\omega^m$ is the gauge topological current ($\partial_n\omega^n
=F_{mn}^{(a)}\widetilde{F}^{mn}_{(a)}$, where $a$ denote indices
of the adjoint representation).

The Lagrangian can be continued to Euclidean space by following
the usual prescriptions \cite{a}. In particular, the vector field $h^m$
is replaced by $ih^m$ due to its dependence on
the totally antisymmetric $\epsilon$ tensor. The bosonic
part of the Lagrangian becomes
\begin{equation}
{\cal L_B}~=~\frac{1}{2\al}{\cal R}+\frac{1}{4\al
l^2}\partial_m l\,\partial^m l
+\frac{1}{4\al l^2}h_m h^m +\frac{k}{4l}F_{mn}^{(a)}F^{mn}_{(a)}+
G_{i\bar{\jmath}}D_m z^i D^m\bar{z}^{\bar{\jmath}},
\end{equation}
where $z^i$ are scalar components of chiral superfields
and $D_m$ are gauge covariant derivatives.
We will consider nonperturbative effects due to gauge fields
which do not couple to these scalars -- a pure gauge
``hidden'' sector in superstring terminology.
We also neglect auxiliary
components of vector supermultiplets which vanish for supersymmetric
backgrounds.

We are interested in the solutions of classical field equations
which generalize the usual self-dual (or anti-self-dual) gauge
configurations. Due to the dilaton-dependence of gauge kinetic terms,
such solutions must necessarily involve both gauge fields and
a space-time dependent dilaton.
The configurations
$$
F_{mn}^{(a)}=\pm\widetilde{F}_{mn}^{(a)}\eqno{(3a)}$$
$$ h_m=\pm\partial_ml \eqno{(3b)}$$ \addtocounter{equation}{+1}
do indeed satisfy all field equations in a flat vierbein background
$e_{\mu}^m= \delta_{\mu}^m$, with constant (possibly zero) VEVs of
scalars $z^i$. Specific solutions of eqs.(3) have already
appeared in the context of superstring solitons
\cite{chs,rey,s}. In particular,
the one-instanton configuration  of size $\rho$,
centered at position $x_0$,
\begin{equation}
A_m^{(a)}=2\,\frac{\eta_{amn}(x-x_0)^n}{(x-x_0)^2+\rho^2}\, ,
\end{equation}
where $\eta_{amn}$ is the 't Hooft symbol \cite{th},
is accompanied by the dilaton
configuration \cite{s}
\begin{equation}
l(x)~=~l_0+4\al k\,\frac{(x-x_0)^2+2\rho^2}{[(x-x_0)^2+\rho^2]^2}\, .
\end{equation}

The Lagrangian $\cal L_B$ becomes a total divergence when
evaluated on a self-dual (and anti-self-dual)  solution of eqs.(3).
The instanton action
\begin{equation}
S~=~\int\!d^4x\sqrt{g}\,{\cal L_B}~=~\frac{8\pi^2 k}{l_0}\label{s}
\end{equation}
depends on the asymptotic value $l_0$ of the dilaton only. This is exactly
the same action as for a single instanton configuration in a
Yang-Mills system without a dilaton,
with the gauge coupling constant identified with
$\sqrt{l_0/k}$. The instanton size $\rho$ and its position $x_0$
are the bosonic zero modes of this action, in addition to the
usual rotational and gauge zero modes.
The non-perturbative
amplitudes will involve the usual suppression factor of $e^{-S}$.

In order to compute non-perturbative contributions to
correlation functions, we need the fermionic zero
modes in the instanton background. The fermionic field equations following
from the Lagrangian of ref.\cite{adam} involve terms
which are linear and trilinear in fermions. We will argue later on that
the trilinear part vanishes for solutions of the linearized equations.
The linearized field equations in the bosonic background under
consideration are
\begin{equation}
\sigmab^n\partial_n\chi+\frac{3h_n-4\partial_nl}{4l}
\sigmab^n\chi+i\al kF^{mn}_{(a)}\sigmab_{mn}\lambdab^{(a)}
-\frac{1}{2}(h_n+\partial_nl)\sigmab^m\sigma^n\psib_m=0 \label{e1}
\end{equation}
\begin{equation}
\sigma^nD_n\lambdab^{(a)}+\frac{h_n-2\partial_nl}{4l}
\sigma^n\lambdab^{(a)}
+\frac{i}{2l}F_{mn}^{(a)}\sigma^{mn}\chi
+\frac{1}{2}[F_{mn}^{(a)}+\widetilde{F}_{mn}^{(a)}]\sigma^m\psib^n
=0\label{e2}
\end{equation}
\begin{equation}
\epsilon^{mnpq}\sigma_n\partial_p\psib_q+
\frac{h_n-\partial_nl}{4l^2}\sigma_n\sigmab_m\chi
+\frac{\al k}{2l}[F^{mn}_{(a)}-\widetilde{F}^{mn}_{(a)}]
\sigma_n\lambdab^{(a)}=0 \label{e3}
\end{equation}
where $\chi$ is the dilatino, $\lambdab^{(a)}$ are the gauginos, and
$\psib_n$ are the gravitinos.
In the equations above, we neglected
other fermions which can be set zero for constant (possibly zero)
VEVs of their scalar superpartners.
In the equations conjugate to (\ref{e1}-\ref{e3}),
$h_n$ is replaced by $-h_n$ and $\widetilde{F}_{mn}^{(a)}$
by $-\widetilde{F}_{mn}^{(a)}$.

The simplest way to obtain solutions of eqs.(\ref{e1}-\ref{e3}) is to perform
supersymmetry and superconformal transformations
on the bosonic background. The transformation rules are
\begin{equation}
\delta\chi=-\frac{i}{2}(h_n+\partial_nl)\sigma^n\bar{\varepsilon}-2i\,l\zeta,
\label{chi}
\end{equation}\begin{equation}\label{lam}
\delta\lambdab^{(a)}=\frac{1}{2}F_{mn}^{(a)}\sigmab^{mn}\bar{\varepsilon},
\end{equation}\begin{equation}\label{psi}
\delta\psib_n=(\partial_n-\frac{h_n}{4l})\bar{\varepsilon}-\sigmab_n\zeta,
\end{equation}
where $\varepsilon$ and $\zeta$ are the supersymmetry and superconformal
transformation parameters, respectively. The transformation rules for
the conjugate fermions
$\bar{\chi}$, $\lambda^{(a)}$ and $\psi_n$ are obtained by formal complex
conjugation of eqs.(\ref{chi}-\ref{psi}), with $h_n\rightarrow -h_n$.

We begin with a self-dual (one-instanton) configuration
$F_{mn}^{(a)}=\widetilde{F}_{mn}^{(a)}$, $h_m=\partial_ml$.
We choose the supergravity gauge $\sigma^n\psib_n=\sigmab^n\psi_n=0$.
A supersymmetry transformation with $\varepsilon=0$, $\bar{\varepsilon}
=l^{1/4}\bar{\theta}$, where $\bar{\theta}$ is a constant spinor, yields
$$\chi=-il^{1/4}\partial_nl\,\sigma^n\bar{\theta}\eqno{(13a)}$$
$$\lambdab^{(a)}=\half l^{1/4}F_{mn}^{(a)}\sigmab^{mn}\bar{\theta},
\eqno{(13b)}$$
while leaving zero all other fermions, including the gravitino.
This fermion configuration does indeed satisfy eqs.(\ref{e1}-\ref{e3}).
Another solution with a vanishing gravitino field can be generated
by a supersymmetry transformation with $\varepsilon=0$,
$\bar{\varepsilon}=l^{1/4}x_n\sigmab^n\theta$, where $\theta$ is a constant
spinor, followed by a superconformal transformation with
$\zeta=l^{1/4}\theta$, $\bar{\zeta}=0$. This solution is
not normalizable though with respect to any reasonable norm, since
$\chi(x)\rightarrow -2i\,l_0^{\,5/4}\theta$ at $x\rightarrow\infty$.
A normalizable solution can be obtained by subtracting
$\chi_0(x)=-2i\,l_0\,l^{1/4}\theta$, a trivial solution of
eqs.(\ref{e1}-\ref{e3}) with all other fermions set zero, which can be thought
of as the fermionic superpartner of the constant dilaton mode $l_0$.
The final form of the corresponding normalizable solution is
$$\chi=-i\,l^{1/4}x_n\partial_ml\,\sigma^m\sigmab^n\theta
-2i\,l^{1/4}(l-l_0)\theta\eqno{(14a)}$$
$$\lambdab^{(a)}=-l^{1/4}x^mF_{mn}^{(a)}\sigmab^n\theta\eqno{(14b)}$$
\addtocounter{equation}{+2}
with all other fermions equal to zero.

The fermionic field configurations of eqs.(13) and (14) satisfy
not only the linearized eqs.(\ref{e1}-\ref{e3}), but also the full equations
which involve additional trilinear terms. The reason is that the Lagrangian
is invariant under global $U(1)$ chiral symmetry transformations,
with charge +1 fermions
$\chi$, $\lambdab^{(a)}$ and $\psib_n$, and charge $-$1 conjugate fermions.
The trilinear terms in fermionic field equations have charges
+1 or $-$1, hence they always involve some fermions of opposite charges
whereas in solutions (13) and (14)
all non-vanishing fermions
have the same charges. Eqs.(13) and (14) provide therefore
4 fermionic zero modes of the action in a self-dual instanton background.
The zero modes in an anti-instanton background
are obtained by formal complex conjugation of the instanton modes.
If the gauge group is larger than $SU(2)$, additional zero modes
can be constructed by following the usual procedures of
instanton calculations. Without losing generality, we
restrict our attention to the minimal case of
$SU(2)$, with 8 real bosonic zero modes and 4 fermionic zero
modes given by eqs.(13) and (14).

The space-time-dependent part of the dilaton field, the antisymmetric
tensor field and the dilatino component of the fermionic zero modes
are all of order  ${\cal O}(\al )$. The integration measure
over the zero modes may also contain corrections of order
${\cal O}(\al/\rho^2 )$ to the standard instanton measure with the coupling
constant $g=\sqrt{l_0/k}$.
These corrections are not important though for the following computations
of non-perturbative amplitudes, where we will limit ourselves
to extracting the leading order behaviour in $\al$.
The supersymmetric $SU(2)$ instanton zero mode measure is
\begin{equation}
d\mu \,e^{-S}~=~ \kappa\,\Lambda^6g^{-4} d\rho^2\, d^4x_0\,l_0^{-1} d^2\theta
\,d^2\bar{\theta}\label{dmu}
\end{equation}
where $\kappa$ is a known constant \cite{a}, and
the classical factor $e^{-S}$ has been incorporated into the measure.
In eq.(\ref{dmu}), $g$ has been identified with the gauge coupling
constant at the energy scale $(\al)^{-1/2}$, and $\Lambda
=(\al)^{-1/2}\exp(-4\pi^2/3g^2)$ is the strong interaction scale.

In order for a correlation function to receive a non-vanishing
instanton contribution, it must contain at least 4 fermions
which can saturate the fermionic zero modes of eqs.(13) and (14).
As an example, we compute the following two-point
correlation functions:
\begin{eqnarray}
\langle\lambdab^2(x)\lambdab^2(y)\rangle &\!\equiv\!&
\langle\lambdab^{(a)}(x)\lambdab_{(a)}(x)\;
\lambdab^{(a)}(y)\lambdab_{(a)}(y)\rangle,\\
\langle\lambdab^2(x)\chi^2(y)\rangle &\!\equiv\!&
\langle\lambda^{(a)}(x)\lambda_{(a)}(x)\; \chi(y)\chi(y)\rangle.
\end{eqnarray}
Only the zero mode components contribute to these
amplitudes, giving
\begin{eqnarray}
\hspace*{-5mm}\langle\lambdab^2(x)\lambdab^2(y)\rangle &\!\!=\!\!&
\int\! d\mu \,e^{-S} l^{1/2}(x)
F^2(x)\,(x-y)^2 \,l^{1/2}(y)F^2(y)\,
\frac{\theta^2\bar{\theta}^2}{16} , \\[3mm]
\hspace*{-5mm}\langle\lambdab^2(x)\chi^2(y)\rangle &\!\!=\!\!&
\int\! d\mu \,e^{-S} l^{1/2}(x)
F^2(x)\,(x-y)^{-2} l^{1/2}(y)\,
\{\frac{\partial}{\partial y}[\,l(y)-l_0\,](x-y)^2\}^2
\,\frac{\theta^2\bar{\theta}^2}{4}.
\end{eqnarray}
As a result of the zero mode integration we obtain
\begin{eqnarray}
\langle\lambdab^2(x)\lambdab^2(y)\rangle &\!=\!&
\frac{(64\pi)^2}{20}\kappa\,\Lambda^6g^{-4}\,
\{\,1+{\cal O}[\frac{\al}{(x-y)^2}]\,\}\, ,\label{l4}\\[2mm]
\langle\lambdab^2(x)\chi^2(y)\rangle &\!=\!&
\frac{191(64\pi)^2}{75}\kappa\,\Lambda^6g^{-4}\al k^2\,\frac{\al}{(x-y)^2}\,
\{\,1+{\cal O}[\frac{\al}{(x-y)^2}]\,\}\, . \label{l2h2}
\end{eqnarray}
All zero mode integrations are convergent,
as in the globally supersymmetric case \cite{a}.

Similar computations can be performed for other correlation functions.
The common property of all these instanton-induced amplitudes
is their non-vanishing $U(1)$ charge equal +4 (or $-$4 for anti-instantons);
the $U(1)$ symmetry is broken down to its discrete
${\sf Z\hspace{-1.5mm}Z}_{4}$
subgroup. For gauge groups larger than $SU(2)$ the unbroken
subgroup is $\sf Z\hspace{-1.5mm}Z_{\it n}$, where $n$ is the number of
fermionic zero modes (\eg $n=2M$ for $SU(M)$, $n=60$ for $E_8$ \etc).
Instantons do {\em not} contribute non-zero VEVs for the
$U(1)$-invariant four-fermion operators present in the Lagrangian,
nor do they induce a scalar potential. Local supersymmetry is
preserved by instanton solutions.

The superstring-type supergravity under consideration becomes
a standard globally supersymmetric theory in the limit
$\al\rightarrow 0$. The dilaton $l(x)$ is frozen then at its VEV $l_0$.
Out of all instanton-induced amplitudes, only the
pure gaugino correlation function of eq.(\ref{l4}) remains non-zero
in this limit.
The result agrees with ref.\cite{a}. As expected,
the supergravity corrections are negligible at distances $(x-y)^2\gg\al$.

We are in a position now to discuss the relation between
instanton-induced amplitudes and a gaugino condensate
$\langle\lambdab^2 (x)\rangle$. In the limit $(x-y)^2\rightarrow\infty$
the gaugino amplitude (\ref{l4}) goes to a constant,
as expected on the basis of unbroken supersymmetry \cite{a}. One
can argue now in the spirit of ref.\cite{a} that
the amplitude factorizes in this limit into $\langle\lambdab^2 (x)\rangle
\langle\lambdab^2 (y)\rangle$. Self-consistency of the theory,
more precisely factorization and clustering, imply then the
existence of yet another non-perturbative effect that gives rise
to gaugino condensation -- instantons have too many
fermionic zero modes to produce a two-gaugino condensate.
Note that this effect should not produce a dilatino condensate,
as seen from the large $(x-y)^2$ behaviour of the amplitude (\ref{l2h2}).
In pure supersymmetric Yang-Mills theory, supersymmetry is
protected by Witten's index theorem \cite{w}.
This may be different though in the context of supergravity.
The full supergravity transformation
of the dilatino contains a term of the form given in eq.(\ref{chi}),
with $\zeta (x)=\al k \,l^{-1}\lambdab^2(x)\,\varepsilon(x)/8$, where
$\varepsilon(x)$ is the local supersymmetry transformation parameter.
Gaugino condensation $\langle\lambdab^2 (x)\rangle\neq 0$
gives rise to a constant term of order
${\cal O}(\al )$ in the supergravity transformation
for the dilatino. One can argue that local
supersymmetry is dynamically
broken, with the dilatino identified as the goldstino \cite{fgn}.

Although our analysis has been restricted so far to a
supergravity theory with the gauge kinetic terms depending
on the dilaton in a way dictated by superstring theory in the tree
approximation, our strategy can be readily extended to
more complicated cases. It can be shown that self-dual solutions
satisfying conditions similar to eqs.(3) exist for a
supergravity theory defined by the d-term of $f(\widehat{L})\, e^{-\al G/2}$,
where $f$ is an arbitrary real function. They also exist in the presence
of the so-called Green-Schwarz term \cite{gs} which appears at the one-loop
level as a result of integrating out the heavy string states \cite{agnt}.
In all these cases the zero modes can be constructed in a very
similar way, and the calculation of non-perturbative amplitudes
is straightforward.

The results of this work are not surprising.
The fact that supersymmetry is gauged in supergravity theory
turns out to be without much importance for instantons.
They preserve supersymmetry while breaking a chiral
$U(1)$ symmetry. Non-perturbative amplitudes receive corrections
and some new correlation functions appear. However, all of them
are suppressed at large distances.
Hence the presence of instantons
does not trigger dynamical supersymmetry breaking in supergravity theories,
although such a breaking
may occur as a result of other non-perturbative effects.

SJR benefitted from useful discussions with C.G. Callan and J. Distler, and
acknowledges
hospitality of the Northeastern University at early stage of this work.
TRT acknowledges hospitality of the Ecole Polytechnique where
this work was completed. TRT is grateful to
I. Antoniadis, L. Alvarez-Gaum\'e, C. Bachas, Y. Meurice, and especially
to G. Veneziano, for very helpful discussions and suggestions.
\newpage\setlength{\parsep}{0pt}\setlength{\itemsep}{0pt}

\end{document}